\documentclass[10pt,twocolumn,aps,prc,showpacs,superscriptaddress,preprintnumbers,floatfix,nofootinbib,longbibliography]{revtex4-1}
\usepackage{multirow}
\usepackage{hhline}
\usepackage{bbm}
\usepackage{graphics}  % needed for figures
\usepackage{graphicx}  % needed for figur\es
\usepackage{dcolumn}   % needed for some tables
\usepackage{bm}        % for math
\usepackage{amssymb}   % for math
\usepackage{amsmath}   % for math
\usepackage{amsfonts}
\usepackage[percent]{overpic}
\usepackage{natbib}
\usepackage{xcolor}
\usepackage{tikz}
\usepackage[colorlinks=true,linkcolor=blue,citecolor=blue,urlcolor=blue]{hyperref}

\newcommand{\npart}{\ensuremath{N_{\mathrm{part}}}}
\newcommand{\mnpart}{\ensuremath{\langle N_{\mathrm{part}}\rangle}}
\newcommand{\sqrtsNN}{\ensuremath{\sqrt{s_{\mathrm{NN}}}}}
\newcommand{\pT}{\ensuremath{p_{\mathrm{T}}}}
\newcommand{\mpT}{\ensuremath{\langle p_{\mathrm{T}}\rangle}}
\newcommand{\dmpT}{\ensuremath{\langle\langle p_{\mathrm{T}}\rangle\rangle}}

\newcommand{\cTwo}{\ensuremath{\langle c_{2}\rangle}}

\newcommand{\auau}{Au+Au }

\definecolor{lime}{HTML}{A6CE39}
\DeclareRobustCommand{\orcidicon}{
	\begin{tikzpicture}
	\draw[lime, fill=lime] (0,0) 
	circle [radius=0.16] 
	node[white] {{\fontfamily{qag}\selectfont \tiny ID}};
	\draw[white, fill=white] (-0.0625,0.095) 
	circle [radius=0.007];
	\end{tikzpicture}
	\hspace{-2mm}
}
\foreach \x in {A, ..., Z}{%
	\expandafter\xdef\csname orcid\x\endcsname{\noexpand\href{https://orcid.org/\csname orcidauthor\x\endcsname}{\noexpand\orcidicon}}
}
\foreach \x in {A, ..., Z}{%
	\expandafter\xdef\csname orcid\x\endcsname{\noexpand\href{https://orcid.org/\csname orcidauthor\x\endcsname}{\noexpand\orcidicon}}
}

\begin{document}
%\linenumbers

\title{Mean-\texorpdfstring{\pT}{pT} fluctuations in \auau{} collisions at \sqrtsNN{} = 3.0--19.6~GeV within JAM2}

\newcommand{\htu}{Henan Normal University, Xinxiang, 453007, China}
\newcommand{\hinst}{Institute of Nuclear Science and Technology, Henan Academy of Sciences, Zhengzhou, 450046, China}
\newcommand{\moe}{Key Laboratory of Nuclear Physics and Ion-beam Application (MOE), and Institute of Modern Physics, Fudan University, Shanghai 200433, China}
\newcommand{\fudan}{Shanghai Research Center for Theoretical Nuclear Physics, NSFC and Fudan University, Shanghai 200438, China}
\author{Yaqi Liu\orcidD{}}\affiliation{\htu}\affiliation{\hinst}
\author{Liuyao Zhang\orcidA{}}\email{zhangly@hnas.ac.cn}\affiliation{\hinst}
\author{Chunjian Zhang\orcidC{}}\email{chunjianzhang@fudan.edu.cn}\affiliation{\moe}\affiliation{\fudan}
\author{Jinhui Chen\orcidB{}}\email{chenjinhui@fudan.edu.cn}\affiliation{\moe}\affiliation{\fudan}
\author{Chunwang Ma}\affiliation{\hinst}\affiliation{\htu}

\begin{abstract}
Event-by-event mean-\pT{} fluctuations probe initial-state fluctuations and their evolution through the dynamics of heavy-ion collisions. 
We study second-order mean-\pT{} fluctuations in
\auau{} collisions at \sqrtsNN{} = 3.0--19.6~GeV using JAM2 in the RQMDv mean-field mode with the MH2 parameterization. The model qualitatively 
reproduces the measured identified-particle \pT{} spectra, providing a single-particle baseline for the fluctuation analysis.
The scaled fluctuation $k_2$ decreases with increasing \mnpart{} and shows broad agreement
with the available measurements at 7.7--19.6~GeV, whereas its calculated centrality dependence
is stronger than that in the data at 3.0--4.5~GeV. For the combined proton-plus-antiproton sample, the unnormalized correlator \cTwo{} is positive
and larger than that for charged pions. 
The charged-pion correlator \cTwo{} is negative or consistent with zero over most centrality intervals at 3.0 and 3.5 GeV, becomes weakly positive at 4.5~GeV, 
and remains positive at higher energies. Its energy evolution resembles the change in the reaction-plane elliptic flow, but this
comparison does not establish a common microscopic origin.
These calculations provide species-dependent predictions within a transport model without
an explicit partonic stage.  Isolating the contributions of mean fields, rescattering, and spectator interactions requires controlled variations of the transport dynamics. 
\end{abstract}
\maketitle

\section{Introduction} 
\label{sec:introduction}

Event-by-event fluctuations of the mean transverse momentum, \mpT{}, provide a sensitive probe of fluctuations in the
initial energy deposition and transverse size, and of their conversion into collective dynamics in relativistic heavy-ion collisions~\cite{Stodolsky:1995ds,Shuryak:1997yj,Gavin:2003cb,Broniowski:2009fm,Gavin:2011gr,Bozek:2012fw,Jia:2021qyu}.
In a hydrodynamic description, at comparable initial entropy, a more compact initial system develops stronger transverse pressure gradients and consequently 
a larger radial-flow response, leading to an anticorrelation between the initial transverse size and the final-state
\mpT{}~\cite{Broniowski:2009fm, Bozek:2012fw}. Because this response depends on the equation of state (EoS) and transport properties
of the produced matter, mean-\pT{} fluctuations and the related differential radial-flow observable $v_0(\pT)$~\cite{ALICE:2025iud,ATLAS:2025ztg,Jia:2025rab,Wan:2025rzg} 
provide information complementary to anisotropic-flow 
observables. More generally, mean-\pT{} observables have been used to extract thermodynamic properties of hot QCD matter~\cite{Gardim:2019xjs, Cao:2021zhy,Mu:2026uaz}.
This sensitivity is particularly relevant in the RHIC Beam Energy Scan (BES) region, where the system 
reaches large net-baryon densities and its evolution is influenced by baryon stopping, hadronic interactions, 
nuclear mean fields, and the finite-density QCD EoS~\cite{Bzdak:2019pkr,Chen:2024aom,Nara:2021fuu,Parida:2026kyo}.

The dynamical component of mean-\pT{} fluctuations is commonly quantified using multiparticle \pT{} correlators, 
which eliminate self-correlations and suppress the statistical contribution associated with finite particle multiplicity~\cite{PHENIX:2002aqz,STAR:2005vxr,ALICE:2014gvd, Bhatta:2021qfk,Chen:2025vwl}. Measurements from SPS to LHC energies 
have revealed finite two-particle \pT{} correlations whose magnitude generally decreases with increasing multiplicity or system size, 
consistent with the dilution of correlated particle pairs as the number of particle-emitting sources increases~\cite{NA49:1999inh,CERES:2008wlj,PHENIX:2003ccl,STAR:2003cbv,STAR:2013sov, ALICE:2014gvd,ALICE:2023tej,ATLAS:2024jvf,ALICE:2024apz}. 
Recently, the STAR Collaboration reported measurements of two-particle \pT{} correlations for midrapidity 
charged particles in fixed-target \auau{} collisions at \sqrtsNN{} = 3.0--7.7~GeV~\cite{STAR:2026vjv}. 
The scaled fluctuation approximately follows the independent-source expectation, $1/\sqrt{\npart{}}$, over a broad centrality 
range. However, a clear breakdown of this scaling is observed in central collisions, together with a statistically 
significant non-monotonic beam-energy dependence. 
A recent PNJL study has discussed a possible connection between a dip in the temperature cumulant and this non-monotonic \pT{} trend~\cite{Liu:2026pro}.
These observations motivate quantitative tests of noncritical
low-energy dynamics~\cite{Reichert:2026ztj} before isolating effects associated with critical dynamics~\cite{Stephanov:1999zu}. 
In the same energy region, identified-hadron elliptic-flow measurements exhibit a rapid change near 4.5~GeV, which has been 
interpreted as evidence for the onset of dominant partonic interactions~\cite{STAR:2025owm}.

Several theoretical frameworks have been employed to investigate the origin of mean-\pT{} fluctuations. 
Hydrodynamic calculations relate these fluctuations to event-by-event variations in the initial energy deposition and 
transverse size, which are converted into fluctuations of the collective transverse expansion~\cite{Bozek:2012fw, Giacalone:2020lbm, Chatterjee:2017mhc, Schenke:2020uqq,Parida:2024ckk}. 
Correlations between mean-\pT{} and anisotropic flow further provide additional sensitivity to the initial nuclear geometry and 
its dynamical response, and have recently been used to constrain nuclear deformation~\cite{Giacalone:2021uhj,Giacalone:2021udy, 
Jia:2021tzt, Magdy:2021ocp, Wang:2024mce,STAR:2024wgy}.
HIJING supplies a reference for source superposition and correlations without collective
final-state evolution~\cite{Wang:1991hta, Bhatta:2021qfk,Wang:2026hak}, 
while AMPT has been used to study the centrality, beam-energy, and particle-species dependence of mean-\pT{} 
fluctuations~\cite{Zhang:1999bd,Lin:2004en,Xu:2020pxj, Zhang:2025yyd}. In particular, an AMPT study already covers 3.0--19.6~GeV
and examines higher-order correlators, particle classes, and subevent selections~\cite{Zhang:2025yyd}. 
Hadronic transport calculations with UrQMD provide a complementary noncritical reference
including hadronic production, decays, and rescattering~\cite{Bass:1998ca, Bleicher:1999xi, STAR:2019dow}. At the lowest BES
energy, baryon stopping, spectator-induced shadowing, and nuclear mean-field interactions become increasingly 
important. Therefore, a microscopic transport calculation incorporating these effects is required to establish an 
appropriate baseline for mean-\pT{} fluctuations in baryon-rich matter~\cite{Nara:1999dz, Zhang:2018wlk, Nara:2021fuu}.

The question addressed here is how second-order \pT{} correlations, particularly their particle-species dependence,
behave in a transport calculation with baryonic mean fields but no explicit partonic stage. We study second-order mean-\pT{} fluctuations in \auau{} collisions at \sqrtsNN{} = 3.0--19.6~GeV using 
JAM2 in the RQMDv mode. We examine the single-particle description through identified-particle spectra and
inclusive mean transverse momenta reconstructed from those spectra. We then investigate the centrality and beam-energy dependence of
the two-particle correlator \cTwo{} and its scaled fluctuation $k_2$, with particular emphasis on
comparisons with recent STAR fixed-target measurements. The particle-species dependence is examined 
separately for charged pions and the combined proton-plus-antiproton sample. Finally, the reaction-plane elliptic flow 
provides context for the evolution of transverse dynamics.
The comparison concerns energy-dependent trends, rather than a measured event-by-event correlation
between the two observables. With a single mean-field parameterization, the present study
establishes a model reference rather than an extraction of the EoS or a separation of individual mechanisms. 

\section{Methodology and model setup}
\label{sec:Methodology_JAM} 
\subsection{Mean-\pT{} fluctuation observable}

For an event $e$ containing $N_e$ selected particles, the event-averaged mean transverse momentum is defined as
\begin{equation}
\dmpT
=
\left\langle
\frac{1}{N_e}
\sum_{i=1}^{N_e}p_{\mathrm{T},i}
\right\rangle_{\mathrm{ev}},
\label{eq:meanpt}
\end{equation}
where $\langle\cdots\rangle_{\mathrm{ev}}$ denotes an average over events. In the present generator-level analysis, no detector-efficiency weights are applied. Following the multiparticle \pT{}-correlator formalism~\cite{Bhatta:2021qfk,Jia:2017hbm,Zhang:2025yyd}, we define
$\delta p_{\mathrm{T},i}=p_{\mathrm{T},i}-\dmpT$. The second-order correlator is
\begin{equation}
\cTwo{}
=
\left\langle
\frac{\sum_{i\neq j}
\delta p_{\mathrm{T},i}\delta p_{\mathrm{T},j}}
{N_e(N_e-1)}
\right\rangle_{\mathrm{ev}},
\label{eq:c2}
\end{equation}
where the sum runs over ordered pairs of distinct particles. The condition $i\neq j$ removes self-correlations, and
only events with $N_e\geq2$ contribute. 
A nonzero \cTwo{} can receive contributions from event-by-event fluctuations in particle production, collective expansion, resonance 
decays, conservation laws, string fragmentation, and hadronic interactions~\cite{Voloshin:1999yf,Bhatta:2021qfk,STAR:2019dow,STAR:2026vjv}.

To reduce the dependence on the overall transverse-momentum scale and facilitate comparisons among collision energies and
centralities, we use the dimensionless scaled fluctuation
\begin{equation}
k_2 =
\frac{\sqrt{\cTwo{}}}{\dmpT}, 
\qquad \cTwo{}>0.
\label{eq:k2}
\end{equation}
The observable $k_2$ is evaluated for the inclusive charged particles, for which \cTwo{} remains positive within 
the investigated kinematic range.

\begin{figure*}[htb]
\centering
\includegraphics[scale=0.82]{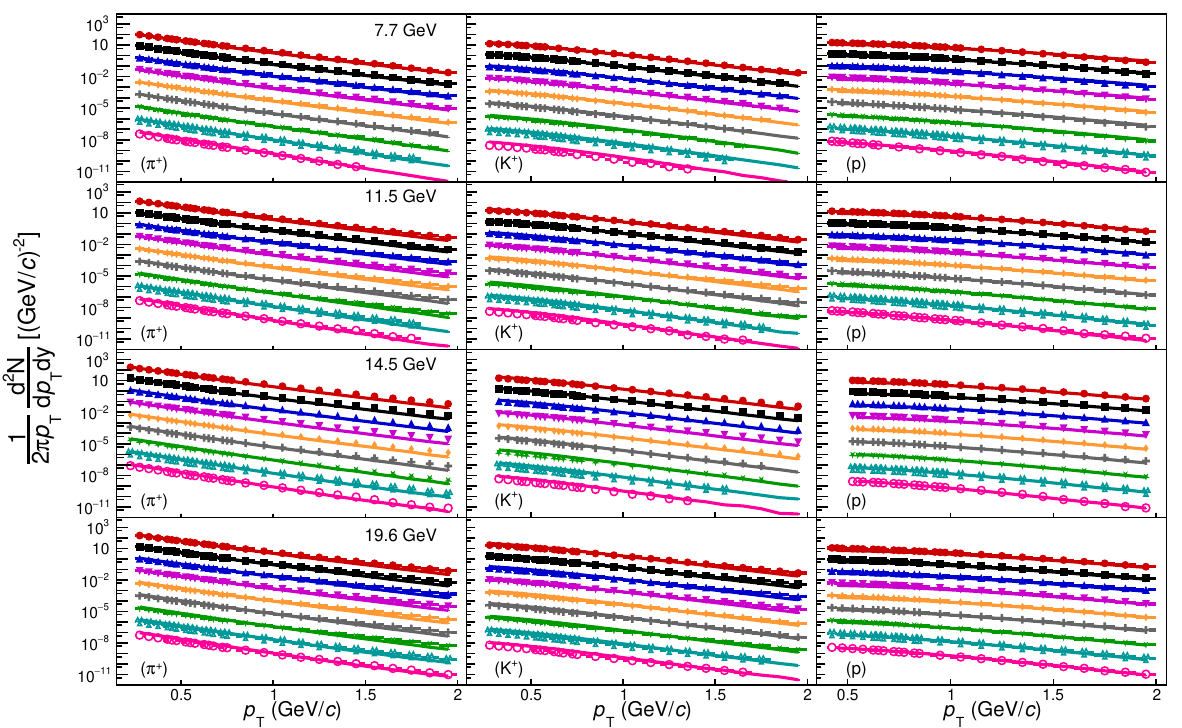}
\caption{Transverse-momentum spectra of $\pi^{+}$ (left), $K^{+}$ (middle), and protons (right) at $|y|<0.1$ 
         in \auau{} collisions at \sqrtsNN{} = 7.7, 11.5, 14.5, and 19.6~GeV. Symbols denote STAR measurements~\cite{STAR:2017sal, STAR:2019vcp}, 
         and curves show the JAM2/RQMDv calculations. For visibility, the 0--5\% spectra are shown
         without rescaling ($10^{0}$), while the spectra for successive centrality intervals are 
         multiplied by $10^{-1}$, $10^{-2}$, and so forth.}
\label{fig:pTSpec}
\end{figure*}

\subsection{JAM2 model and event selection}

The collision dynamics are simulated using the JAM2 microscopic transport model~\cite{Nara:1999dz,Nara:2021fuu}. 
JAM2 describes the nonequilibrium hadronic evolution through binary hadron-hadron scattering, 
resonance excitations and decays, and string excitations and fragmentation.
At BES energies, the system can reach high net-baryon densities, where nuclear mean fields
and their associated pressure response can significantly affect the collective dynamics~\cite{Nara:2020ztb}. 

The calculations employ the RQMDv mode of JAM2, in which the Skyrme-type density- and momentum-dependent hadronic mean fields are
implemented as Lorentz-vector potentials~\cite{Nara:2021fuu}. In RQMDv, the scalar potential vanishes,
$S_i=0$, and the scalar effective mass is $m_i^*=m_i$; this does not remove the
momentum dependence of the single-particle dispersion relation. The kinetic four-momentum and the corresponding
on-shell constraint are 
\begin{equation}
p_i^{*\mu}=p_i^\mu-V_i^\mu,
\qquad
p_i^{*2}-m_i^2
=
(p_i-V_i)^2-m_i^2
=0.
\label{eq:rqmdv}
\end{equation}
The density-dependent part of the mean field is parameterized by a Skyrme-type potential,

\begin{equation}
U_{\mathrm{sk}}(\rho)
=
\alpha\left(\frac{\rho}{\rho_0}\right)
+
\beta\left(\frac{\rho}{\rho_0}\right)^{\gamma},
\label{eq:skyrme}
\end{equation}
where $\rho$ denotes the baryon density entering the mean-field prescription and $\rho_0=0.168~\mathrm{fm}^{-3}$ is the nuclear saturation density. 
This expression specifies the density dependence of the interaction, not the full vector
potential $V_i^\mu$ or the complete EoS. The MH2 parameterization additionally contains a two-range,
momentum-dependent vector interaction constructed with Lorentzian kernels from the local phase-space distribution~\cite{Nara:2021fuu}. 

We employ the momentum-dependent hard EoS parameterization MH2, with a nuclear incompressibility of $K=380~\mathrm{MeV}$
at saturation density. This incompressibility characterizes cold symmetric nuclear matter
at saturation and does not by itself determine the finite-temperature, high-density pressure. The same model parameters are used at all collision energies and centralities, 
and the evolution is followed up to $50~\mathrm{fm}/c$. The \pT{} correlations are evaluated for charged particles within
$|\eta|<0.5$ and $0.2<\pT{}<2.0~\mathrm{GeV}/c$. Identified-particle results are presented separately for $\pi^{\pm}$ and $p+\bar{p}$. 
Charged-kaon correlators are not reported because of the limited statistics.

The centrality percentiles are assigned according to the \npart{} distribution obtained directly from the JAM2 event record.  
For the fluctuation analysis, events are classified into 0--10\%, 10--20\%, 20--30\%, 30--40\%, 40--50\%, 50--60\%, 60--70\%, and 70--80\% intervals, 
each characterized by its corresponding \mnpart{}. This truth-level classification differs from an experimental multiplicity-based centrality
selection; equal values of \mnpart{} need not imply identical event ensembles. The reaction-plane elliptic flow is calculated for
$|\eta|<1.0$ and $0.2<\pT{}<2.0~\mathrm{GeV}/c$ as
\begin{equation}
v_2
=
\left\langle
\cos\left[2\left(\phi-\Psi_{\mathrm{RP}}\right)\right]
\right\rangle ,
\label{eq:v2rp}
\end{equation}
where the average is taken over the selected particles and events. The impact-parameter direction defines the reaction plane and is fixed at
$\Psi_{\mathrm{RP}}=0$ in the generated events. Consequently, $v_2=\langle\cos(2\phi)\rangle$. This quantity refers to the known reaction plane and is not automatically identical to
an experimental event-plane or multiparticle-cumulant flow estimate.

\section{Results} 
\label{sec:results}

Figure~\ref{fig:pTSpec} compares the JAM2/RQMDv calculations with the measured \pT{} spectra of $\pi^{+}$,
$K^{+}$, and protons at midrapidity in \auau{} collisions at \sqrtsNN{} = 7.7--19.6~GeV. The calculations capture 
the centrality ordering of the spectra and qualitatively describe their shapes over the measured \pT{} range. This level
of agreement tests the average single-particle distributions at 7.7--19.6~GeV, but does not
by itself validate two-particle correlations or the single-particle description below 7.7~GeV.

\begin{figure}[htb]
\centering
\includegraphics[scale=0.45]{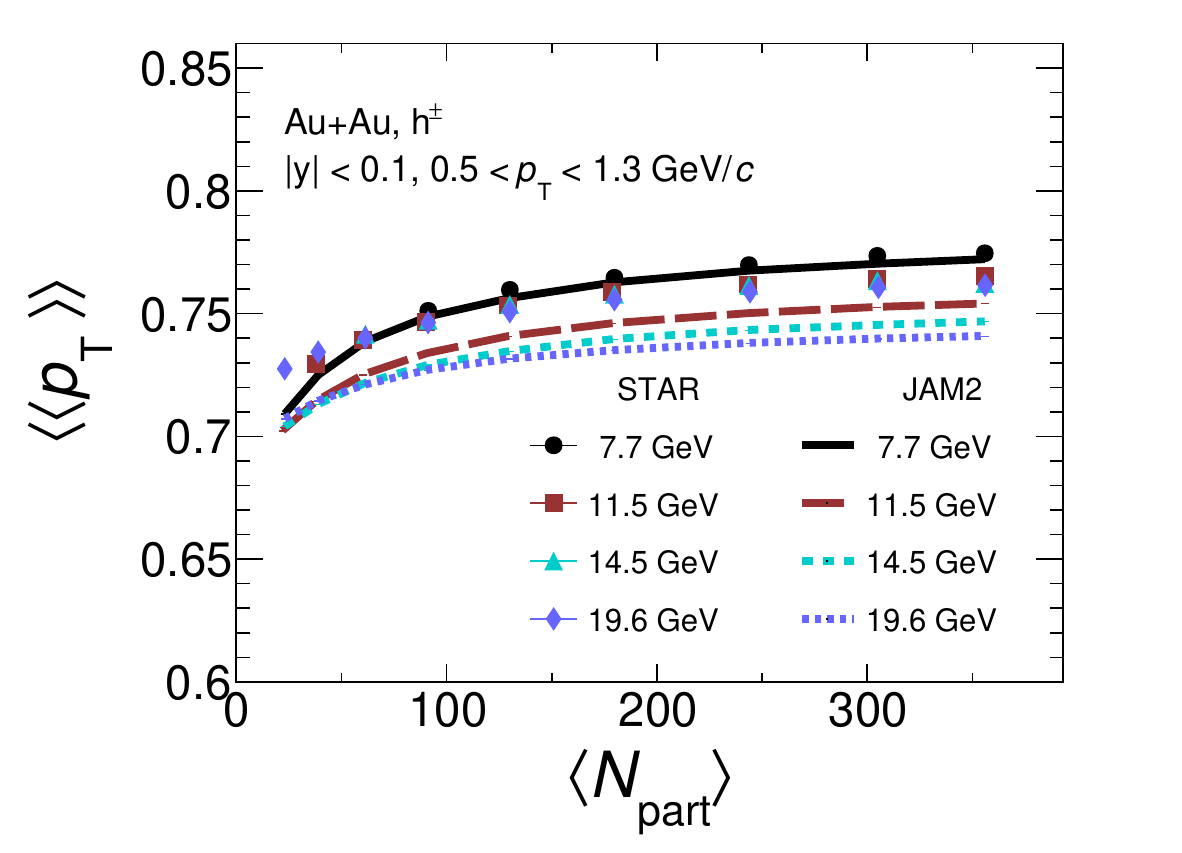}
\caption{Inclusive charged-hadron  mean transverse momentum as a function of \mnpart{} in \auau{} collisions at 
         \sqrtsNN{} = 7.7--19.6~GeV. Curves show the JAM2 results, and symbols denote values reconstructed 
         from STAR identified-particle spectra~\cite{STAR:2017sal, STAR:2019vcp}. The common acceptance
         is $|y|<0.1$ and $0.5<\pT{}<1.3~\mathrm{GeV}/c$.}
\label{fig:mpT_Npart}
\end{figure}

Figure~\ref{fig:mpT_Npart} compares the inclusive charged-hadron mean transverse momentum from JAM2 with values reconstructed from 
the STAR identified-particle spectra.
Both the JAM2 results and the reconstructed values increase from peripheral to central collisions. 
The model curve is close
to the reconstructed values at 7.7~GeV and lies below them at 11.5--19.6~GeV, while preserving
their beam-energy ordering. 
Within the restricted acceptance, the higher mean at lower energy can be influenced by
the increasing proton fraction~\cite{STAR:2017sal}; separating this composition effect from
changes in the species-dependent spectra requires a yield-weighted decomposition.

\begin{figure}[htb]
\centering
\includegraphics[scale=0.42]{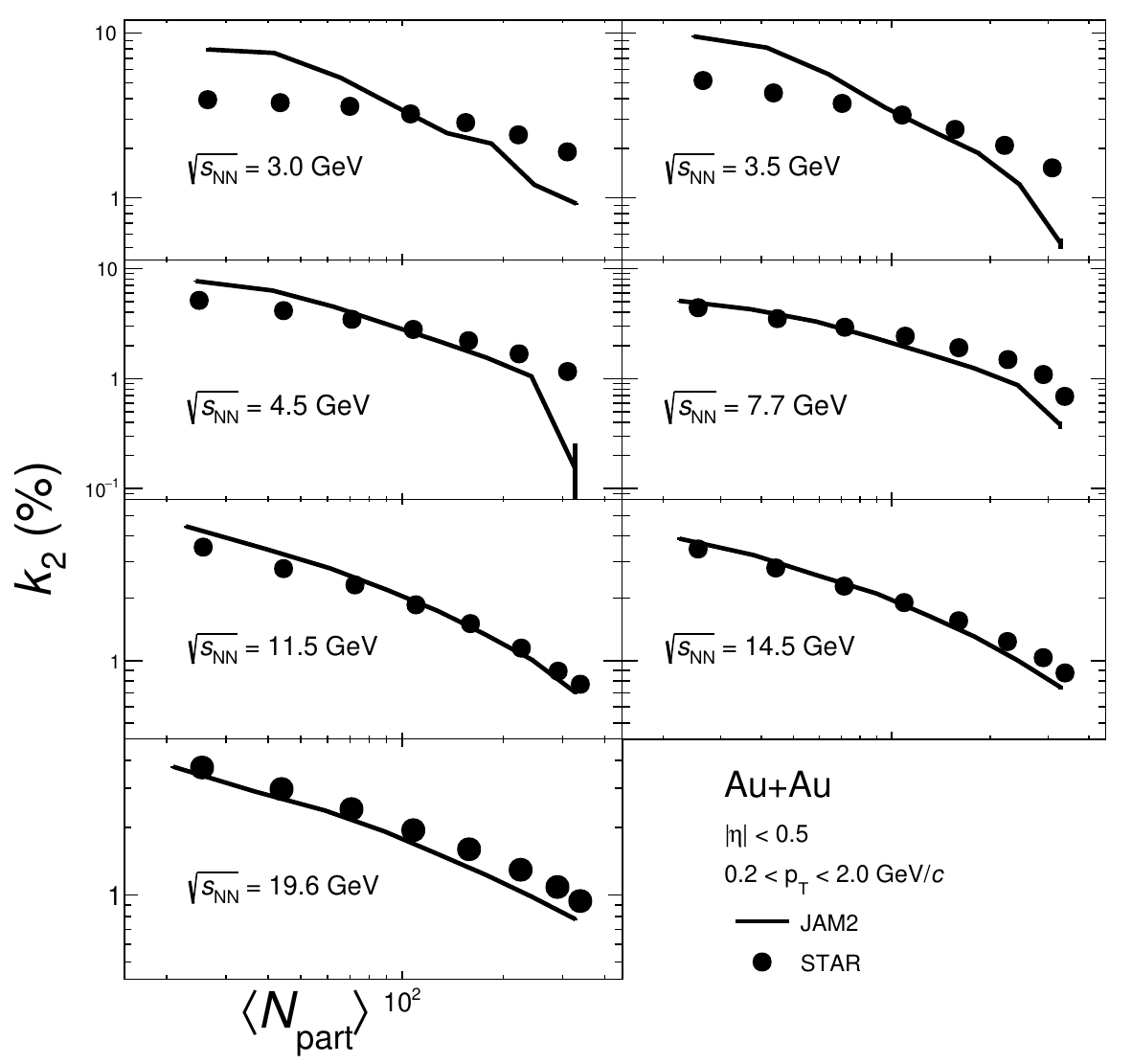}
\caption{Scaled second-order mean-\pT{} fluctuation $k_2$, expressed as a percentage,
         as a function of \mnpart{} in \auau{} collisions at \sqrtsNN{} = 3.0--19.6~GeV. 
         Curves show the JAM2/RQMDv results, and symbols denote the available STAR measurements~\cite{STAR:2019dow,STAR:2026vjv}.
         Charged particles are selected within $|\eta|<0.5$ and $0.2<\pT{}<2.0~\mathrm{GeV}/c$.}
\label{fig:k2_Npart}
\end{figure}

Figure~\ref{fig:k2_Npart} presents the scaled second-order mean-\pT{} fluctuation $k_2$ as a function of \mnpart{}. 
At all investigated energies, $k_2$ decreases with increasing \mnpart{}, broadly consistent with the dilution of
correlated particle pairs as the number of particle-emitting sources increases. A monotonic decrease alone does not establish the independent-source scaling
$k_2\propto\mnpart^{-1/2}$, which additionally assumes approximately unchanged source properties
and a source number proportional to \npart{}. The calculation shows broad agreement with the magnitude and centrality dependence of
the available STAR measurements at \sqrtsNN{} = 7.7--19.6~GeV. 
At 3.0--4.5~GeV, however, the calculated $k_2$ exhibits a stronger dependence on \mnpart{} than observed in the data, 
particularly toward central collisions. This difference identifies a limitation of the present
comparison. Its dynamical origin cannot be isolated without evaluating the effects of centrality
selection, acceptance, and particle composition alongside variations of the transport settings.

\begin{figure*}[htb]
\centering
\includegraphics[scale=0.85]{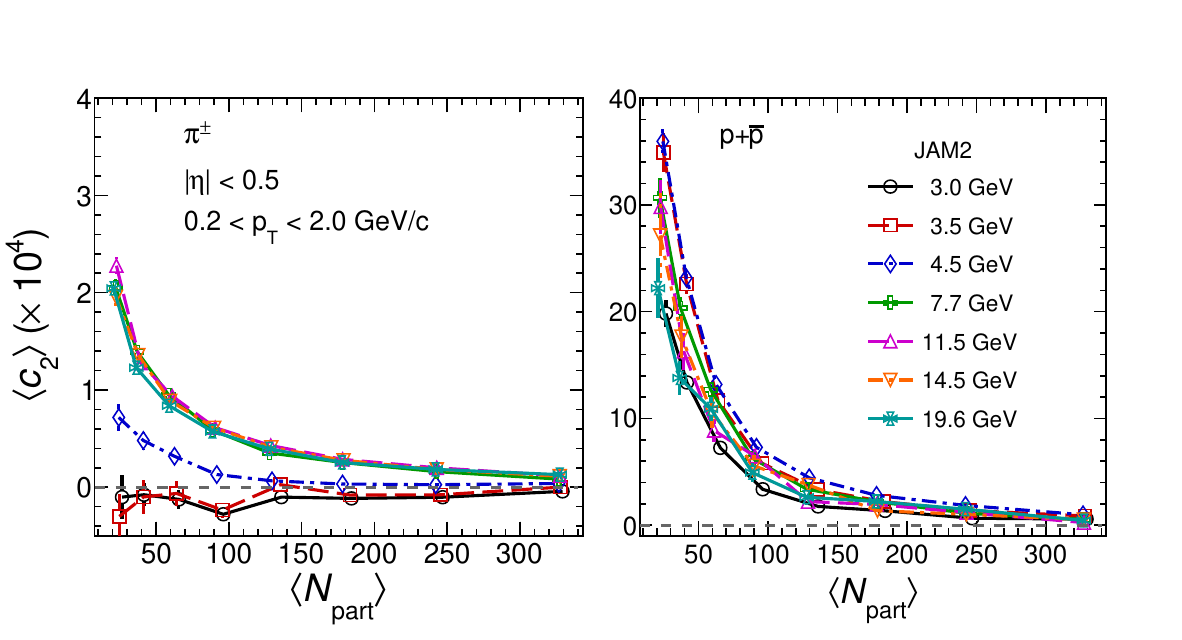}
\caption{Unnormalized two-particle correlator \cTwo{}, in $(\mathrm{GeV}/c)^2$, as a function of
         \mnpart{} for the combined $\pi^++\pi^-$ sample (left) and the combined $p+\bar p$
         sample (right) in \auau{} collisions at \sqrtsNN{} = 3.0--19.6~GeV,
         calculated with JAM2/RQMDv--MH2. Both same-species and cross-species pairs enter
         each combined sample. Particles are selected in $|\eta|<0.5$ and
         $0.2<\pT{}<2.0~\mathrm{GeV}/c$.}
\label{fig:c2_pid}
\end{figure*}

The particle-species dependence of the unnormalized correlator \cTwo{} is shown in Fig.~\ref{fig:c2_pid}. Each result uses the mean of its own selected particle sample. The combined samples contain
both same-species and cross-species pairs; the $p+\bar p$ result is not a separate
proton--antiproton cross-correlation. 
For the combined proton-plus-antiproton sample,  \cTwo{} remains positive throughout the investigated energy and centrality
ranges and decreases rapidly with increasing \mnpart{}. Its magnitude is substantially
larger than that for charged pions, particularly in peripheral collisions. Since \cTwo{} is not normalized 
by the species-dependent mean transverse momentum, its larger magnitude for protons and antiprotons may partly 
reflect their larger characteristic transverse-momentum scale. 
Differences in multiplicity,
the distribution of pair types, and particle composition can also contribute~\cite{Reichert:2026ztj}.
The unnormalized hierarchy therefore does not by itself establish stronger relative fluctuations. 
Fluctuations in baryon stopping, radial expansion, and baryonic mean-field dynamics may also contribute, 
although their individual effects are not separated in the present calculation. 

The charged-pion correlator exhibits a qualitatively different beam-energy dependence. 
It is negative or compatible with zero over most centrality intervals at 3.0 and 3.5~GeV, becomes 
weakly positive at 4.5~GeV and remains positive with a modest energy dependence at \sqrtsNN{} = 7.7--19.6~GeV.
A negative \cTwo{} denotes an anticorrelation of the scalar transverse momenta of distinct
pions relative to the mean of the selected pion ensemble; it is not a negative variance.
Conservation constraints and resonance decays can affect momentum correlations~\cite{Borghini:2006yk,Stephanov:1999zu},
but neither vector transverse-momentum conservation nor decay kinematics fixes the sign
of this scalar correlator in a restricted acceptance.
Their contributions must be evaluated using the same particle selection and estimator.
Event-by-event changes in a common transverse-momentum scale can generate positive correlations,
irrespective of the sign of the azimuthal anisotropy. Identifying the origin of the pion trend
requires charge-resolved correlations, uncertainty estimates near zero, and controlled model comparisons.

\begin{figure}[htb]
\centering
\includegraphics[scale=0.45]{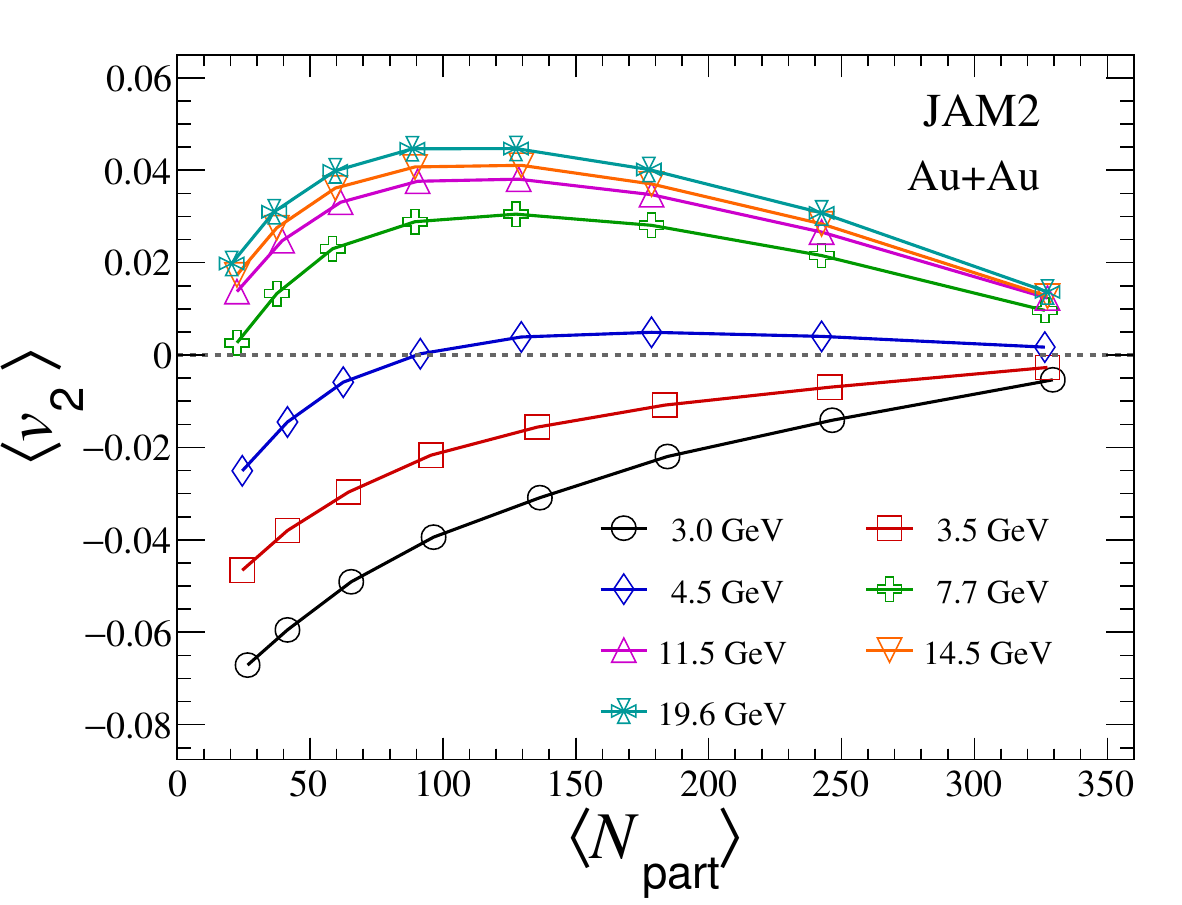}
\caption{Reaction-plane elliptic flow $v_2 =\langle\cos[2(\phi-\Psi_{\mathrm{RP}})]\rangle$ as a function of \mnpart{}
         in \auau{} collisions at \sqrtsNN{} = 3.0--19.6~GeV calculated with JAM2/RQMDv. The reaction-plane angle is fixed at 
         $\Psi_{\mathrm{RP}}=0$. Particles are selected within $|\eta|<1.0$ and $0.2<\pT{}<2.0~\mathrm{GeV}/c$.}
\label{fig:v2_energy}
\end{figure}

To examine the accompanying evolution of collective dynamics, Fig.~\ref{fig:v2_energy} presents the reaction-plane elliptic flow ($v_2$)
calculated with JAM2. At \sqrtsNN{} = 3.0 and 3.5~GeV, $v_2$ remains negative over the full centrality range, consistent with
spectator-induced out-of-plane squeeze-out~\cite{E895:1999ldn}. At 4.5~GeV, $v_2$ changes from negative values in peripheral 
collisions to small positive values for $\mnpart{}\gtrsim100$. At 7.7--19.6~GeV, $v_2$ remains positive and reaches a maximum in midcentral collisions,
indicating the increasing dominance of in-plane expansion.

The beam-energy evolution of $v_2$ qualitatively resembles that of the charged-pion \cTwo{}.
The sampled energies, however, do not determine a unique common transition energy.
The comparison is limited by the different observables and pseudorapidity acceptances;
a species-matched comparison is also required. In particular, rotating all transverse momenta
by $\pi/2$ relative to a fixed reaction plane changes the sign of $v_2$ while leaving \cTwo{}
unchanged. Thus the observed sign patterns are not connected by an algebraic relation.
STAR has interpreted the emergence of
constituent-quark-number scaling of identified-hadron $v_2$ as evidence for the onset of dominant partonic interactions by
\sqrtsNN{} = 4.5~GeV~\cite{STAR:2025owm}. The present integrated reaction-plane calculation
does not test that scaling and does not confirm or exclude the experimental interpretation.
JAM studies have found an important spectator-shadowing effect on low-\pT{}
meson directed flow~\cite{Liu:2024ugr}, while comparisons of
hadronic and partonic transport at 3.0 and 4.5~GeV have examined
constituent-quark-number scaling of identified-hadron $v_2$~\cite{Liu:2025eml}.
Recent theoretical work also examines how spectator shadowing can modify constituent-quark-number
scaling in an idealized source model~\cite{Reichert:2026dsb}; it does not predict the
pion \pT{} correlator studied here.
Our calculation establishes that the reported qualitative trends occur in a model without an
explicit partonic stage. It does not separate spectator, rescattering, and mean-field contributions.

\section{Discussion and Summary}
\label{sec:discussion}

We have studied second-order mean-\pT{} fluctuations in \auau{} collisions at
\sqrtsNN{} = 3.0--19.6~GeV using JAM2/RQMDv with a common MH2 mean-field parameterization.
The identified-particle spectra provide a qualitative single-particle check at 7.7--19.6~GeV;
they do not independently validate the two-particle predictions or the lower-energy spectra.
The inclusive scaled fluctuation decreases toward central collisions and broadly follows
the available data at 7.7--19.6~GeV. At 3.0--4.5~GeV its calculated centrality dependence
is stronger than that observed. This difference warrants matched centrality and acceptance
tests before it is assigned to a particular missing dynamical mechanism.

The species dependence provides a more differential prediction. The combined $p+\bar p$
sample has a positive unnormalized correlator, larger than that of charged pions, but this
hierarchy alone does not imply stronger relative fluctuations. For charged pions, the correlator
is negative or consistent with zero over most centrality intervals at 3.0 and 3.5~GeV,
becomes weakly positive at 4.5~GeV, and remains positive at higher energies.
Its qualitative parallel with reaction-plane elliptic flow does not establish a common origin:
the scalar \pT{} correlator and the azimuthal harmonic probe distinct aspects of the dynamics.
The present results neither test constituent-quark-number scaling nor establish critical behavior.

The calculation supplies a specified transport baseline without an explicit partonic stage.
It does not isolate the effects of spectator interactions, rescattering, or mean fields,
and a single MH2 setting cannot provide a quantitative EoS constraint.
The next tests are charge-resolved and signed normalized correlators, checks of centrality
and decay contributions, and controlled comparisons of cascade and alternative mean-field settings.
An energy scan at fixed centrality, with denser coverage between 3 and 5~GeV, would directly
address whether the model reproduces the low-energy structure of the inclusive correlation.
Extending the calculation below 3~GeV and implementing the appropriate experimental acceptance
would also be necessary for quantitative predictions for CEE at HIRFL-CSR~\cite{Wang:2026CEE}; comparisons with
the HIAF program likewise require its relevant beam energies and acceptance~\cite{Yang:2013yeb}.

\section{Acknowledgements} 
We thank Yasushi Nara and Weijie Fu for insightful discussions and
Chen Zhong for providing a stimulating research environment. This work was supported in part by
the National Key Research and Development Program of China under Contract Nos. 2022YFA1604900 and 2024YFA1612600; 
the State Key Laboratory of Heavy Ion Science and Technology, Institute of Modern Physics, Chinese Academy of Sciences Nos. HIST2026CO19;
Henan High-End Foreign Expert Recruitment Program Nos. HNGD2026049; institutional research projects of the Henan Academy of Sciences
Nos. 20251846001 and 20260646001; the National Natural Science Foundation of China (NSFC) under Contract Nos. 12025501 and 12547102;
the Natural Science Foundation of Shanghai under Contract No. 23JC1400200.
%\end{CJK*}

\bibliography{reference}

\end{document}